\documentclass[
 aip,
 amsmath,amssymb,
 reprint,
]{revtex4-1}

\usepackage{graphicx}
\usepackage{dcolumn}
\usepackage{bm}

\usepackage[utf8]{inputenc}
\usepackage[T1]{fontenc}
\usepackage{mathptmx}
\usepackage{etoolbox}
\usepackage{xcolor}
\makeatletter
\def\@email#1#2{
 \endgroup
 \patchcmd{\titleblock@produce}
  {\frontmatter@RRAPformat}
  {\frontmatter@RRAPformat{\produce@RRAP{*#1\href{mailto:#2}{#2}}}\frontmatter@RRAPformat}
  {}{}
}

\makeatother

\begin{document}

\title{Optimization of laser-driven proton acceleration in a near-critical-density plasma}

\author{Guanqi Qiu}
 \affiliation{State Key Laboratory of Nuclear Physics and Technology, and Key Laboratory of HEDP of the Ministry of Education, CAPT, School of Physics, Peking University, Beijing 100871, China}

\author{Qianyi Ma}
 \affiliation{State Key Laboratory of Nuclear Physics and Technology, and Key Laboratory of HEDP of the Ministry of Education, CAPT, School of Physics, Peking University, Beijing 100871, China}

\author{Deji Liu}
 \affiliation{State Key Laboratory of Nuclear Physics and Technology, and Key Laboratory of HEDP of the Ministry of Education, CAPT, School of Physics, Peking University, Beijing 100871, China}

\author{Dongchi Cai}
 \affiliation{State Key Laboratory of Nuclear Physics and Technology, and Key Laboratory of HEDP of the Ministry of Education, CAPT, School of Physics, Peking University, Beijing 100871, China}

\author{Zheng Gong}
\affiliation{Institute of Theoretical Physics, Chinese Academy of Sciences, Bejing 100190, China}

\author{Yinren Shou}
 \affiliation{Institute of Modern Physics, Fudan University, 200433 Shanghai, China}

\author{Jinqing Yu$^{*}$}
 \email{jinqing.yu@hnu.edu.cn}
 \affiliation{Hunan Provincial Key Laboratory of High-Energy Scale Physics and Applications, School of Physics and Electronics, Hunan University, Changsha 410082, China}
 
\author{Xueqing Yan$^{*}$}
 \email{x.yan@pku.edu.cn}
 \affiliation{State Key Laboratory of Nuclear Physics and Technology, and Key Laboratory of HEDP of the Ministry of Education, CAPT, School of Physics, Peking University, Beijing 100871, China}
 \affiliation{Beijing Laser Acceleration Innovation Center, Huairou, Beijing, 101400, China}
 \affiliation{Institute of Guangdong Laser Plasma Technology, Baiyun, Guangzhou, 510540, China}

\date{\today}

\begin{abstract}
Optimizing laser and plasma parameters is crucial for enhancing accelerated proton energy in laser-driven proton acceleration with finite laser energy for applications such as cancer therapy. Tight focusing plays a significant role in improving laser-driven proton acceleration, which is generally believed as a result of the enhancement of laser intensity. However, we find that even at a fixed laser intensity, reducing the focal spot size still enhances the proton energy. Through particle-in-cell simulations and theoretical modeling, we find that at a small spot size (0.8 $\mu$m), the maximum proton energy is enhanced by 56.3\% compared to that obtained at a conventional spot size (3 $\mu$m). This improvement is attributed to the dominance of ponderomotive-force-driven electrons at reduced spot sizes, which generate stronger charge-separation fields that propagate at higher velocities. Furthermore, to optimize proton acceleration, we analytically derive an ideal plasma density profile that promotes phase-stable proton acceleration, yielding an additional energy increase of 61.3\% over the case of a tightly focused laser interacting with a planar target of uniform density. These findings remain robust under parameter variations, indicating that advanced focusing techniques combined with optimized plasma profiles could relax the demand for high laser energies, thereby reducing the reliance on large-scale laser facilities in medical and scientific applications.

\end{abstract}

\maketitle

\section{Introduction}

Laser-driven plasma acceleration has emerged as a novel acceleration scheme in recent decades \cite{1,2,3,4}, attracting extensive interdisciplinary interest due to its broad application prospects in both fundamental research and practical fields, including nuclear physics \cite{nuclear,nuclear2}, quantum electrodynamics \cite{Reichwein_2025,PhysRevE.102.053212}, materials science \cite{material,material2}, radiography \cite{radiography,radiography2} and radiotherapy \cite{therapy,therapy0,PhysRevResearch.5.L012038}. Since the invention of laser wakefield acceleration (LWFA) in 1979 \cite{tajima1979laser}, it has advanced rapidly for the generation of monoenergetic and low divergence electrons, showing its potential in light sources \cite{albert2016applications,wang2021free}, electron radiography \cite{wan2022direct,wan2023femtosecond,wan2024real,schumaker2013ultrafast,bruhaug2023single} and electron radiotherapy \cite{labate2020toward,guo2025preclinical}. As ions are much heavier than electrons, it is difficult for the wake field to accelerate ions, so other mechanisms are required for ion acceleration. Over the past 20 years, laser-driven proton acceleration has advanced rapidly, giving rise to a series of acceleration mechanisms such as Target Normal Sheath Acceleration (TNSA) \cite{tnsa1,tnsa2}, Radiation Pressure Acceleration (RPA) \cite{esirkepov2004highly,rpa1,rpa2,PhysRevLett.105.065002,hb2,zhang2007efficient,macchi2009light}, Collisionless Shock Acceleration (CSA) \cite{silva2004proton,dollar2012finite,fiuza2012weibel,haberberger2012collisionless,PhysRevLett.119.164801,xiao} and Magnetic Vortex Acceleration (MVA) \cite{bulanov2005ion,bulanov2010generation,bulanov2022advanced}. With current laser technology, proton energy up to 150 MeV has been achieved from the interaction of intense lasers with solid targets \cite{150mev,shou2025laser}, while Au ions have reached energies exceeding 1 GeV~\cite{187,martin2024narrow}. For advanced acceleration mechanisms, protons have been accelerated to over 90 MeV by a novel scheme associated with sheet crossing \cite{shou2025proton,gong2020proton}. Theoretically, GeV-level proton energies are predicted to be attainable at 10$^{22}$ W/cm$^2$ laser intensity~\cite{PhysRevLett.102.145002,PhysRevLett.103.024801,bulanov2010generation}. Moreover, laser driven wakefield proton acceleration represents a promising pathway for tens-of-GeV proton generation\cite{yi2013proton,fp}. Recently, laser-driven proton acceleration in near-critical density (NCD) plasma targets \cite{ncd1,ncd2,ncd3,PhysRevResearch.4.L042031} has gained attention as a promising approach for further energy enhancement. When an ultra-intense laser interacts with NCD plasma, relativistic motion increases the electron mass and reduces the plasma frequency. Once the plasma frequency drops below the laser frequency, the laser can penetrate the target, known as relativistic induced transparency (RIT) \cite{rit1,rit2,rit3}. During this penetration, electrons experience enhanced heating via mechanisms such as direct laser acceleration (DLA) \cite{dla1,dla2}, which finally improves proton energy. 

\begin{figure*}
    \centering
    \includegraphics[width=0.7\textwidth]{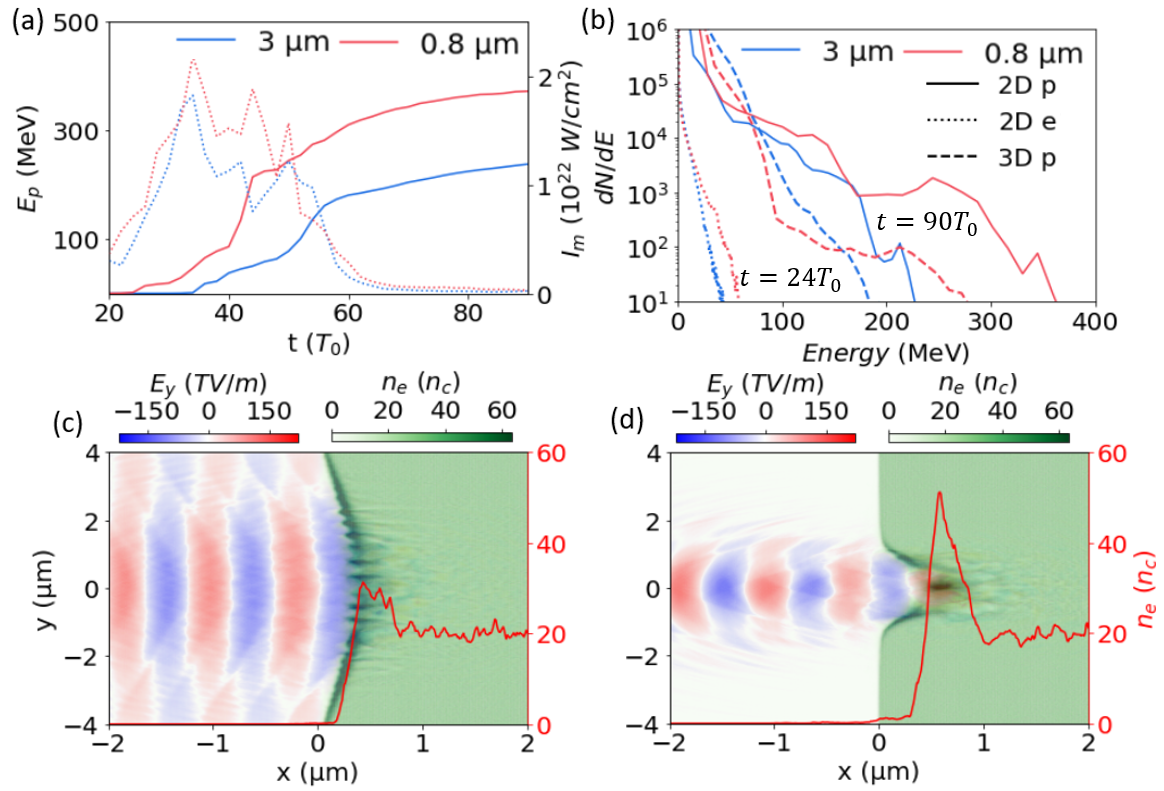}
    \caption{(a) Variation of the energy of the highest-energy proton ($E_p$) with time under focal spot sizes of 3 $\mu$m and 0.8 $\mu$m. The dotted lines show the evolution of the maximum axial laser intensity. (b) Solid lines represent proton energy ($p$) spectrum at the end of  acceleration (t = 90$T_0$), showing significantly higher proton energy with a smaller focal spot compared to the larger one. Dotted lines depict electron energy ($e$) spectrum at t = 24$T_0$, exhibiting higher electron temperature with a smaller focal spot, revealing a faster electron pushing at a smaller focal spot. The dashed lines depict the proton energy spectrum in 3D simulations, confirming the validity of our results. The transverse laser electric field distribution and plasma electron density distribution for focal spot sizes of 3 $\mu$m (c) and 0.8 $\mu$m (d) at t = 24$T_0$, respectively. The red solid lines indicate axial electron density profiles, revealing more pronounced electron accumulation at a smaller focal spot.}
    \label{fig:1}
\end{figure*} 

Despite significant breakthroughs in acceleration mechanisms, current laser energies remain insufficient to accelerate protons beyond the 200 MeV threshold required for applications such as proton radiotherapy\cite{therapy2,therapy3,bulanov2002feasibility}. Given that proton energy generally scales with laser energy \cite{scaling,scaling2,PhysRevE.104.025210}, a critical challenge lies in maximizing proton acceleration efficiency under current laser energy constraints. Thus, optimizing laser and plasma parameters represents a critical avenue for enhancing proton energies, with laser focusing being one promising approach. Recent progress in wavefront correction has enabled near-diffraction-limited focusing, achieving a spot size of 1.1 $\mu$m (full width at half maximum) in experiments \cite{focus}. While tighter focusing is generally expected to boost proton energy through increased laser intensity, the influence of the focal‑spot size itself on energy enhancement has often been neglected. 

In this paper, we demonstrate that reducing the laser focal spots of tightly focused lasers enhances proton energies in laser-near‐critical‑density plasma interactions, even at fixed laser intensity. This implies that higher proton energies can be achieved with lower laser energies. This enhancement originates from the longitudinal laser ponderomotive force, which scales inversely with the square of the laser spot size, becoming significantly stronger at a small focal spot, driving a more efficient electron acceleration. Consequently, a stronger and faster accelerating electric field is generated, leading to higher proton energies. Additionally, we design a down-ramp density profile to achieve velocity matching between protons and the accelerating electric field, further enhancing the proton energy. Through these optimizations, the proton energy can reach near-GeV level, meeting the threshold required for proton radiotherapy.

\section{Results}

A series of two-dimensional (2D) particle-in-cell (PIC) simulations are conducted via relativistic fully self-consistent PIC code EPOCH \cite{epoch}. A p-polarized Gaussian laser pulse with a normalized amplitude of $a_0=eE_L/m_e\omega_L c = 50$, a pulse duration of $\tau$ = 42 fs and a wavelength of $\lambda$ = 800 nm is incident normally on a fully ionized hydrogen target with a density of 20$n_c$ and a thickness of 7.5 $\mu$m, where $n_c=1.7\times10^{21}~cm^{-3}$ is the classical critical density. The simulation domain is a 40 $\mu$m × 10 $\mu$m rectangular box divided into 4000 × 1000 cells, with 10 macro-particles per cell per species. To validate our results, we also conduct three-dimensional (3D) simulations with a 40 $\mu$m × 10 $\mu$m × 10 $\mu$m domain divided into 2000 × 500 × 500 cells and 4 macro-particles per cell per species. To investigate the effect of the laser spot size $\sigma_0$ (defined by the laser field $E_y=E_0exp(-r^2/\sigma_0^2)exp(-t^2/\tau^2)$ at the focus), we vary the laser spot size from 0.8 $\mu$m to 5 $\mu$m, with representative cases of 0.8 $\mu$m and 3 $\mu$m selected for detailed analysis.

\subsection{Optimization by tightly focused laser}

\begin{figure*}
    \centering
    \includegraphics[width=0.7\textwidth]{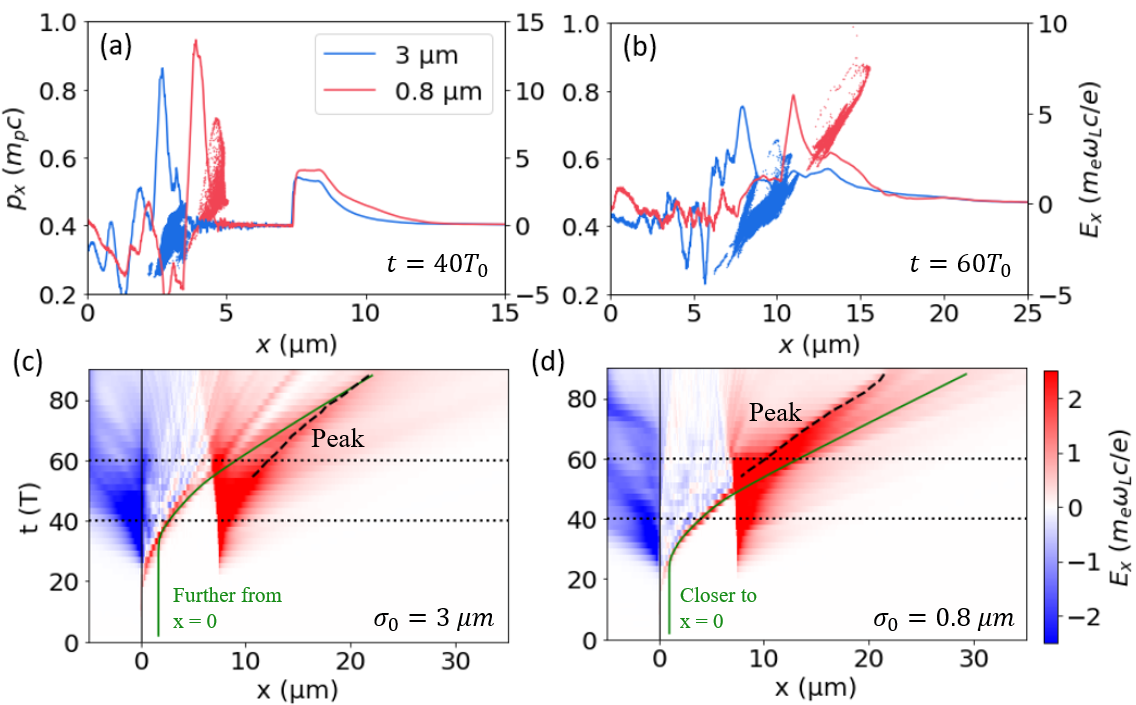}
    \caption{(a) Accelerating electric field and proton phase-space distribution at t = 40$T_0$ (a) and t = 60$T_0$ (b). At t = 40$T_0$, a Hole-Boring field forms inside the target while a TNSA field emerges behind it. By t = 60$T_0$, these two fields have merged and then drift backward. The axial accelerating fields at different times for 3 $\mu$m (c) and 0.8 $\mu$m (d) focal spots, with green lines representing the trajectories of the highest-energy proton. The dotted lines represent t = 40$T_0$ corresponding to (a) and t = 60$T_0$ corresponding to (b). The dashed lines show the location of the accelerating field peak. The black solid lines represent the front surface of the target. The proton accelerated to the highest-energy is from a position more close to the target front surface in (d), showing earlier proton capture by the accelerating field at a smaller focal spot.}
    \label{fig:2}
\end{figure*}

Figure 1(a) shows the temporal evolution of the maximum proton energy for 0.8 $\mu$m and 3 $\mu$m spot sizes, and the proton energy spectrum ($p$) at the end of the acceleration process is shown by the solid curve in Fig.~1(b). It can be observed that when the focal spot is 0.8 $\mu$m, the cutoff energy of protons increases by 56.3\% compared to the case with a 3 $\mu$m spot, reaching 372 MeV in 2D simulations. Although consuming a higher laser energy, the proton energy only reaches 238 MeV in the case with a larger focal spot. This counterintuitive result indicates that reducing the laser spot size, even at a fixed intensity and lower total laser energy, enhances proton acceleration significantly. Considering protons with energies over 50 MeV, we can obtain that for 0.8 $\mu$m spot, the divergence is $\sigma_\theta=\sqrt{\epsilon_{rms}/\beta}$ = 162 mrad (where $\epsilon_{rms}=\sqrt{<y^2><y'^2>-<yy'>^2}$ is the transverse root-mean-square emittance and $\beta=<y^2>/\epsilon_{rms}$ is a Twiss parameter), the proton particle number is $4\times10^{16}$. For 3 $\mu$m spot, the divergence is $\sigma_\theta$ = 171 mrad, the proton particle number is $4\times10^{16}$. In 2D simulations, the sum of proton energies for protons over 50 MeV is 0.64 MJ/m with a spot size of 0.8 $\mu m$ and 0.56 MJ/m with a spot size of 3 $\mu m$. The laser energy is given by $W_L = \int I_0 \exp\left(-2y^2/\sigma_0^2 - 2t^2/\tau^2\right) \, dy \, dt$, where $I_0=5.375\times10^{21}~W/cm^2$ is the laser intensity and $\tau$ = 42 fs is the laser duration. The laser energy is 2.87 MJ/m for $\sigma_0$ = 0.8 $\mu m$ and 10.76 MJ/m with a spot size of $\sigma_0$ = 3 $\mu m$. So the laser-to-proton energy conversion efficiency is 22\% with a spot size of 0.8 $\mu m$ and 5\% with a spot size of 3 $\mu m$. In 3D simulations, the sum of proton energies for protons over 25 MeV is 0.63 J with a spot size of 0.8 $\mu m$ and 1.38 J with a spot size of 3 $\mu m$. The laser energy is given by $W_L = \int I_0 \exp\left(-2(y^2+z^2)/\sigma_0^2 - 2t^2/\tau^2\right) \, dy \, dz \, dt$. The laser energy is 4.07 J for $\sigma_0$ = 0.8 $\mu m$ and 57.2 J with a spot size of $\sigma_0$ = 3 $\mu m$. So the laser-to-proton energy conversion efficiency is 15.6\% with a spot size of 0.8 $\mu m$ and 2.4\% with a spot size of 3 $\mu m$. The 3D results are depicted by the dashed lines in Fig.~1(b), confirming the validity of our findings. The proton cutoff energy in 3D simulations do not decrease much from 2D simulations, implying most of the proton energy is contributed by the Hole-Boring mechanism \cite{hb1,hb2,hb3,hb4}. To elucidate the underlying physics, we analyze the electron density distributions at t = 24$T_0$ for the 3 $\mu$m case in Fig.~1(c) and the 0.8 $\mu$m case in Fig.~1(d), where $T_0$ is the laser period. When the laser interacts with the NCD plasma, the ponderomotive force expels electrons forward and laterally, forming a channel without electrons and a density spike at the laser front, which can be seen in the axial electron density distribution near x = 0.8 $\mu$m in Figs.~1 (c) and 1 (d). In the case with a focal point of 0.8 $\mu$m in Fig.~1(d), a higher peak electron density can be observed, indicating a stronger laser ponderomotive pushing on electrons. The stronger ponderomotive force at the 0.8 $\mu$m spot size results in a more intense electron acceleration, as further evidenced by the electron energy spectrum ($e$) represented by the dotted curves in Fig.~1(b).

When the laser drives the target electrons, due to the development of transverse instabilities \cite{wan2016physical}, a filamentary density distribution forms at the laser-target interaction surface, as shown in Fig. 1(c). Its wavenumber falls with time \cite{wan2020effects}. For a large focal spot, this filamentary density distribution damages the structure of the target surface interacting with the laser, thereby suppressing electron accumulation, which is detrimental to maintaining the charge separation field and proton acceleration. For a small focal spot, the wavelength of the filamentary instability increases during its development, while the transverse extent limits the number of waves, resulting in the structure shown in Fig. 1(d), where only a single filament forms on the axis.

Additionally, as the laser propagates through the plasma, self-focusing occurs. To analyze this, we plotted the temporal evolution of the maximum axial laser intensity, as indicated by the dashed lines in Fig. 1(a). For a larger focal spot, although self-focusing occurs, it takes place near the rear surface of the target and requires a longer propagation distance. By the time the laser has propagated that far, a large amount of its energy has already been absorbed. Consequently, the peak intensity does not increase significantly, as seen in the last peak of the blue dashed line in Fig. 1(a). For a smaller focal spot, even though the focusing distance is shorter, the spot size is already near its limit, making further focusing difficult to achieve. Therefore, the fact that the two dashed lines in Fig. 1(a) are quite similar indicates that self-focusing does not have a significant impact on proton acceleration.

To explain this observation, we propose a theoretical model. The transverse laser electric field follows a Gaussian distribution
\begin{equation}
    E_y=E_0\frac{\sigma_0}{\sigma}e^{-\frac{r^2}{\sigma^2}-\frac{t^2}{\tau^2}},
\end{equation}
where $\sigma=\sigma_0\sqrt{1+(x/Z_R)^2}$, $\sigma_0$ is the spot size, $Z_R=\pi \sigma_0^2/\lambda$ is the Rayleigh length and r is the transverse distance. $E_0=2\times10^{14}~V/m$ is the electric field at the focus at the 0.8 $\mu$m spot and the 3 $\mu$m spot. The longitudinal ponderomotive force at the axis
\begin{equation}
    F_p=-\frac{m_ec^2}{4\gamma}\frac{\partial a_y^2}{\partial x} =\frac{a_0^2m_ec^2}{2\gamma}\frac{x/Z_R^2}{(1+x^2/Z_R^2)^2}
\end{equation}
reaches its maximum value 
\begin{equation}
   F_p=\frac{3\sqrt{3}m_ec^2a_0^2\lambda}{32\pi \gamma \sigma_0^2}
\end{equation}
at $x=Z_R/\sqrt{3}$. The results clearly demonstrate that as the laser spot size decreases, the ponderomotive force, which scales inversely with the square of the spot size, becomes significantly stronger. This enhanced force drives a larger amount of electrons, consistent with the observation in Fig.~1(d) (0.8 $\mu$m spot), where a more significant electron pile-up is observed compared to Fig.~1(c) (3 $\mu$m spot). Additionally, for a large spot size, the ponderomotive force even can not reach its maximum value in the finite-thickness target.

To elucidate the proton acceleration process under different focal spot sizes, we present the accelerating electric field distributions along with the proton phase-space distributions at t = 40$T_0$ and t = 60$T_0$ in Figs.~2(a) and 2(b), respectively. At t = 40$T_0$, the laser has not yet penetrated the target and the accumulated electrons driven by the laser ponderomotive force generate a charge-separation field, known as the Hole-Boring field. For the 0.8 $\mu$m case, a smaller focal spot results in a stronger longitudinal ponderomotive force. Here, electrons are longitudinally pushed before being transversely expelled, leading to enhanced pile-up (Fig.~1(d)) and a larger charge-separation field, as shown in Fig.~2(a). Additionally, a stronger ponderomotive force induces a higher electron energy, causing the electron peak to propagate faster. This is evidenced by the greater displacement from the target front surface of the electron density peak in Fig.~1(d) (smaller spot) compared to Fig.~1(c) (larger spot) at t = 24$T_0$. Consequently, the electric field in Fig.~2(a) exhibits faster propagation and higher amplitude for the smaller spot, efficiently improving the proton acceleration. At this stage, the TNSA field begins to form on the target rear side, as shown in Fig.~2(a). At t = 60$T_0$, the Hole-Boring field has merged with the TNSA field, forming a distinct "peak-plateau" electric field structure (Fig.~2(b)). Protons pre-accelerated by the Hole-Boring field near the target front surface are then injected into the TNSA drifting field, gaining more energy through the subsequent drift acceleration. Protons at a small spot size are injected into the drifting sheath field earlier, before the sheath field decays or drifts away, leading to a higher proton energy.

To further clarify the proton energy disparity, Figs.~2(c) and 2(d) illustrate the proton acceleration process in the Hole-Boring and drifting field for 3 $\mu$m and 0.8 $\mu$m focal spots, respectively, with green lines marking the trajectories of the highest-energy protons. Obviously, the initial of green solid line is closer to the black solid line in Fig.~2(d), showing that the proton accelerated to the highest-energy is from a position closer to the target front surface for the 0.8 $\mu$m focal spot. This indicates that high-energy protons are captured earlier by the accelerating field at a smaller spot, implying a long Hole-Boring acceleration. From the location of the accelerating field peak represented by the black dashed lines, it is obvious that the velocity of the accelerating field at a small spot is larger than at a larger spot, confirming a stronger longitudinal ponderomotive force and faster electrons. As a result, these protons have gained higher energy before the laser penetrates the target. As they are injected into the drifting sheath field before its amplitude decays or it drifts away, protons at a small spot absorb more energy from the drifting field, achieving higher final energies. However, this enhancement occurs only when the spot size is sufficiently small (typically below 2 $\mu$m). For focal spots larger than 3 $\mu$m, where the effect of the ponderomotive force diminishes, proton acceleration reverts to being dominated by the available laser power. As a result, with increasing laser energy, electron heating becomes more intense. Consequently, at larger focal spots of over 3 $\mu$m under the same laser intensity, the proton energy is generally higher when the focal spot is larger (Fig.~3).

\begin{figure}
    \centering
    \includegraphics[width=0.35\textwidth]{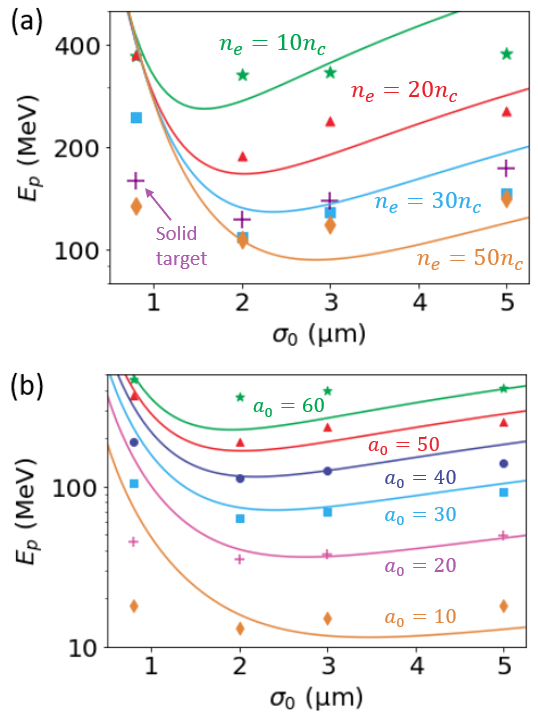}
    \caption{(a) Proton energy as a function of focal spot size under different electron densities. The purple “+” represents the results of simulations with a solid target of 200$n_c$ and 750 nm (the same areal density as the main simulation of NCD target). (b) Proton energy versus focal spot size at varying laser intensities. The theoretical predictions are represented by solid lines and the simulation results are represented by scatters. The laser amplitude $a_0=50$ is used in (a) and the target density $n_e=20n_c$ is used in (b). Across all parameters, proton energy consistently increases as the focal spot size decreases at a spot size smaller than 2 $\mu$m.}
    \label{fig:3}
\end{figure} 

\begin{figure*}
    \centering
    \includegraphics[width=0.7\textwidth]{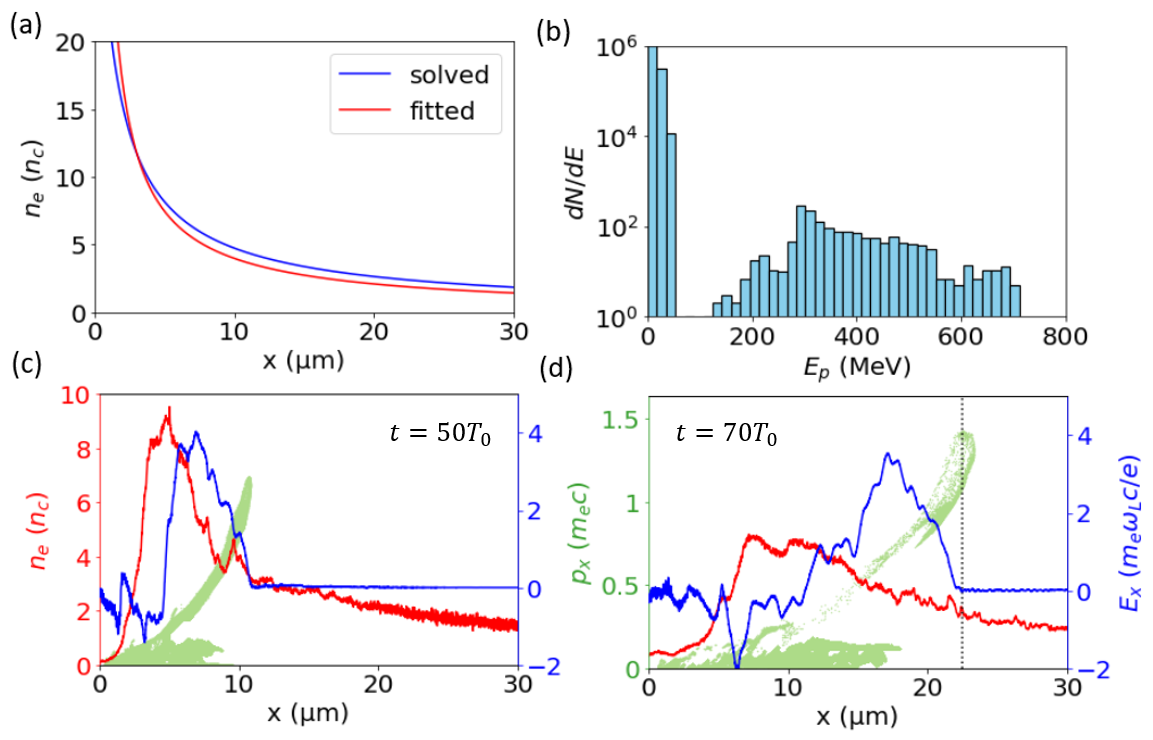}
    \caption{(a) Density profiles derived from numerically solved and function fitted solutions of velocity-matching equation of protons and accelerating field. (b) Proton energy spectrum under this velocity-matching phase-stable acceleration. Electron density distributions, accelerating field distributions, and proton phase-space distributions at t = 50$T_0$ (c) and t = 70$T_0$ (d), demonstrating velocity-matching between protons and the accelerating field. The dotted line in (d) represents the front edge of the accelerating field, illustrating the reflection of protons by the shock.}
    \label{fig:4}
\end{figure*} 

To validate the robustness of these findings, we conducted parameter scans of spot sizes with varying target densities and laser intensities, as shown by scatters in Fig.~3. The results universally demonstrate enhanced proton energies with reduced spot sizes at a spot size smaller than 2 $\mu$m. From Fig. 3(a), this mechanism is confirmed by different laser intensities. From Fig. 3(b), it is obvious that this scheme is robust to a wide range of plasma densities, from near-classical-critic-density target to near-relativistic-critic-density target. Moreover, we also conduct a group of simulations with a solid target of 200$n_c$ and 750 nm, as represented by the purple “+”. Even for solid target, proton energies are also enhanced with reduced spot sizes at a small spot size. However, for $a_0$ > 60, the energy enhancement at a small focal spot turns to be insignificant. This is due to the spot size threshold for small-focus-induced energy enhancement is smaller for higher $a_0$ (higher $\gamma$). A much smaller spot size is required for energy enhancement, but this is restricted by diffraction limit of the laser. Furthermore, we propose a simple theoretical model to describe this phenomenon. The accelerating field is contributed by two distinct electron groups: electrons accelerated by the ponderomotive force at the laser front and electrons accelerated by the DLA mechanism in the channel. The field contributed by the ponderomotive-force accelerated electrons is in direct proportion to the ponderomotive force

\begin{equation}
    E_{x1}=\frac{k_1F_p}{e}=\frac{3\sqrt{3}k_1m_ec^2a_0^2\lambda}{32\pi e\gamma\sigma_0^2},
\end{equation}
where $\gamma=a_0(n_c/n_e)^{0.1}$ is the Lorentz factor and $k_1=16$ is a fit constant. The field contributed by electrons accelerated by the DLA mechanism is in direct proportion to the charge of electrons in the channel $E_{x2}\propto en_cd\pi \sigma_0^2/\epsilon_0$, where d is the target thickness. As the square velocity of the Hole-Boring field is $v_{HB}^2\propto a_0^2n_c/n_e$, \cite{hb2} the drifting field is assumed to have a square velocity $v_d^2\propto a_0^2n_c/n_e$. This assumption is based on the continuity of the accelerating field at the target rear surface in Fig.~2(d). As a result, the proton energy is $E_p\propto v_d^2\propto a_0^2n_c/n_e$. For simplicity, the effect of field velocity on proton energy is considered in the drifting field strength $E_{x2}$, which should be in direct proportion to $a_0^2n_c/n_e$. So, 

\begin{equation}
    E_{x2}= \frac{k_2a_0^2n_c^2ed\pi\sigma_0^2}{\epsilon_0 n_e S}
\end{equation}
where $k_2=0.12\%$ is a fit coefficient to describe the efficiency of accelerated channel electrons and $S=2 \times10^{-5}\sigma _0$ is an area factor corresponding to the spot size. Then we have
\begin{equation}
    eE_x=\frac{3\sqrt{3}k_1m_ec^2a_0^2\lambda}{32\pi \gamma\sigma_0^2}+\frac{k_2a_0^2n_c^2e^2d\pi\sigma_0^2}{\epsilon_0 n_e S}.
\end{equation}

Assuming an acceleration length of $L=10~\mu m$, the proton energy can be obtained as $E_p=eE_xL$. This dual-component framework quantitatively explains the spot-size dependence of proton energy under various parameters. Setting $E_{x1}=E_{x2}$, we can obtain $\sigma_0=(3\sqrt{3} k_1 m_e c^2 \lambda \epsilon_0 n_e S/32\pi^2 e \gamma k_2 n_c^2 ed)^{1/4}$. This is the lower threshold for the laser spot size below which ponderomotive-driven enhancement ceases to be effective. For $a_0=30$ and $n_e=20n_c$, the threshold is $\sigma_0=2.28~\mu m$. The theoretical predictions of our model are represented by solid lines in Fig.~3, which is consistent with simulation results (scatters) in the variation trend. As the model is rough, the coefficients $k_1,k_2$ are only set to fit the trend of spot size effects. $k_1$ is introduced to account for the collective effects of electrons and the role of ions, while $k_2$ compensates for the neglect of transverse effects. In Fig.~2(d), the highest-energy proton has exceeded the drifting field at the end of acceleration, so more optimization is in need to achieve the velocity matching between protons and field.

\subsection{Optimization by down-ramp density target}

Except for laser parameter optimization, proton energies can be further increased through target design, such as employing a density down-ramp \cite{nakamura2010high,wan2019two,reichwein2021robustness}, a technique widely adopted in wakefield electron acceleration to optimize injection \cite{bulanov1998particle,downramp1,downramp2,downramp3}. The down-ramp density profile can be realized through plasma expansion induced by irradiating the rear surface of a target with a long weak laser pulse \cite{downramp,dulat2022subpicosecond}. Due to the greater thickness of the target, the expansion of the rear surface is more significant than that of the front surface \cite{dover2023enhanced}. As a result, a target with a steep front surface and a down-ramp rear surface can be achieved. Alternatively, a stepped descending density profile can be formed by stacking multiple thin NCD targets with progressively decreasing densities. Under laser irradiation, such a stepped density distribution will be smoothed into a continuous down-ramp profile.

During the interaction of lasers with down-ramp near-critical-density targets, phase-stable proton acceleration can be achieved by matching the proton velocity to the accelerating electric field velocity. For linearly polarized lasers, the electric field velocity normalized by the light speed c in the Hole-Boring mechanism is given by \cite{velocity}
\begin{equation}
\beta = \frac{1}{\sqrt{1 + \pi^{2}n_{e}/(n_c a_{0})}},
\end{equation}
where $a_0$ is the normalized laser amplitude and $n_e$ is the electron density. The amplitude of the electric field satisfies \cite{velocity}
\begin{equation}
E_{x} = E_{0}\sqrt{\frac{2(1 - \beta)}{1 + \beta}},
\end{equation}
where $\beta=v/c$ and v is the velocity of the electric field as well as the proton velocity. Velocity matching requires the proton momentum to follow
\begin{equation}
\frac{dp}{dt} = \beta c\frac{dp}{dx} = eE_{x},
\end{equation}
where $p=\gamma m_p c$ and $\gamma = 1/\sqrt{1-\beta^2}$. Solving this we can obtain
\begin{equation}
\frac{1}{4}\ln\left| \frac{1 - \beta}{1 + \beta} \right| + \frac{1}{2(1 - \beta)} - \frac{1}{2} = \frac{\sqrt{2}eE_{0}}{m_{p}c^{2}}x.
\end{equation}
The electron density profile derived from this equation is plotted as the blue curve in Fig.~4(a). The left of Eq.~10 approximates $\beta^2/(2-2\beta)$, and then we get an approximate solution
\begin{equation}
\frac{n_{e}}{n_{c}} = \frac{a_{0}}{2\pi^{2}}\left(\frac{2}{Kx} - 1 + \sqrt{1 + \frac{4}{Kx}}\right),
\end{equation}
as represented by the red curve in Fig.~4(a), where $K=2\sqrt{2} eE_0 /m_p c^2$. The slope of this approximate solution is designed to be steeper than the numerically solved solution, as the laser plasma interaction may result in the expansion of the target and reshape the plasma profile.

To validate this velocity-matching acceleration, we conduct additional simulations to verify the enhancement of proton energy. For a fixed areal density and laser parameters, a 0.8 $\mu$m focus laser is irradiated on a 1.5-$\mu$m-thick target with a uniform electron density of 20$n_c$ following this down-ramp. When a laser irradiates such a target, it first penetrates the uniform density region, exciting a strong electric field that imparts an initial velocity to the protons. Subsequently, the laser begins to interact with the down-ramp density region. As the laser propagates through the decreasing density profile, the electron density peak pushed by the laser radiation moves faster, thereby increasing the speed of the electric field to match the continuously increasing proton velocity. Protons are generated with a cutoff energy exceeding 600 MeV, a 61.3\% increase over the interaction of a 0.8 $\mu$m focus laser with a 7.5-$\mu$m-thick target, as evidenced by the spectrum in Fig.~4(b).

Figures 4(c) and 4(d) illustrate the acceleration process with the proton velocity matching the accelerating field. Protons at the end of the accelerated beam experience stronger fields, continuously overtaking protons in front. This results in a distinct relatively-monoenergetic proton peak (about 50\% energy spread) in Fig.~4(b), in contrast to the exponential spectrum in TNSA. As the field drifts, its velocity increases while the amplitude decays (Fig.~4(d)), maintaining the proton-field velocity matching throughout the acceleration progress. In detail, some high-energy protons exceed the field and leave the acceleration region, as shown in Fig.~4(d). The dotted line in this figure represents the front edge of the accelerating field, demonstrating high-energy protons exceeding the field. These protons are unable to experience the electric field and thus no longer gain energy until the electric field catches up with them again. Then, the field's rising velocity allows it to catch up with these accelerated protons, enabling further proton acceleration.

\section{Conclusion}

In summary, we demonstrate that reducing the laser spot size at fixed intensity enables higher proton energies despite a decrease in total laser energy. This unexpected enhancement is attributed to the electron energy enhancement at small spot sizes, where the ponderomotive force, which scales inversely with the square of the spot size, becomes significantly stronger. This stronger force produces a more intense and faster-propagating charge-separation field, leading to a more efficient proton acceleration. Meanwhile, we design a target with a density down-ramp to achieve velocity matching between protons and the accelerating field, enabling phase‑stable acceleration and a further increase in proton energy. Our results suggest that advances in laser‑focusing techniques can push proton energies to higher levels even with lower total laser energies, thereby reducing the reliance on large‑scale laser facilities. Together with optimized target down‑ramps, spot‑size minimization offers a practical pathway toward efficient laser‑plasma acceleration for applications that demand high particle energies from compact laser systems.

\begin{acknowledgments}
This work was supported by the National Key R\&D Program of China (2025YFF0515103), the Fundamental and Interdisciplinary Disciplines Breakthrough Plan of the Ministry of Education of China (JYB2025XDXIM204), the Strategic Priority Research Program of the Chinese Academy of Sciences (Grant No. XDB1550100), the National Natural Science Foundation of China (Grant No. 11921006) and the National Grand Instrument Project (No. 2019YFF01014400). Deji Liu is supported by the China Postdoctoral Science Foundation (No. 2025M783355). The simulations are supported by the High-Performance Computing Platform of Peking University. This work is partially supported by the National Natural Science Foundation of China (NNSFC) Grants No. 12447101. We acknowledge the HPC Cluster of ITP-CAS for providing computational resources.
\\
\end{acknowledgments}

\bibliography{manuscript}

\end{document}